\documentclass[11pt]{article}
\usepackage{epsf,epsfig,rotating}
\usepackage{amssymb}
\usepackage{amsfonts}
\usepackage{amsmath}
\usepackage{graphicx}

\begin{document}

\title{ \textbf{MSSM inflation and cosmological attractors}}

\author{M.N.~Dubinin\footnote{E-mail: dubinin@theory.sinp.msu.ru}, E.Yu.~Petrova\footnote{E-mail: petrova@theory.sinp.msu.ru}, E.O.~Pozdeeva\footnote{E-mail: pozdeeva@www-hep.sinp.msu.ru}, and S.Yu.~Vernov\footnote{E-mail: svernov@theory.sinp.msu.ru} \\
\small \it  Skobeltsyn Institute of Nuclear Physics, Lomonosov Moscow State University,\\  \small \it Leninskiye Gory~1, 119991, Moscow, Russia}

\date{ \ }

\maketitle

\begin{abstract}
Inflationary scenarios motivated by the Minimal Supersymmetric Standard Model (MSSM) where five scalar fields are non-minimally coupled to gravity are  considered.
The potential of the model and the function of non-minimal coupling are polynomials of two Higgs doublet convolutions. We show that the use of the strong coupling approximation allows to obtain  inflationary parameters in the case when a combination of the four scalar fields plays a role of inflaton. Numerical calculations show that the cosmological evolution leads to inflationary scenarios fully compatible with observational data for different values of the MSSM mixing angle~$\beta$.
\end{abstract}


\section{Introduction}
\label{sec:intro}

The existence of an extremely short
and intense stage of accelerated expansion in the early Universe
(inflation) provides a simple explanation of the astrophysical  data, including the fact that at very large distances the Universe is approximately isotropic, homogeneous and spatially flat~\cite{general1_0, general1_1, general1_2, general1_3}.
The inflationary models are the most reasonable models for the evolution of the early Universe, because they generate the primordial density perturbations finally initiating the formation of galaxies and large-scale structure \cite{Linde:1981mu,general2_0, general2_1, general2_2}.

 The high-precision measurements by the {\it Planck} space telescope show~\cite{PlanckIfl_0,PlanckIfl_1}  that the non-Gaussian perturbations are small. Therefore, the single-field inflationary models are realistic. At the same time  predictions of simplest inflationary models with minimally coupled scalar field include sufficiently large values of $r$, the tensor-to-scalar ratio of density perturbations, and  are  ruled out by the {\it Planck} data. Inflationary scenarios with a minimally coupled scalar field can be improved by adding a tiny non-minimal coupling of the inflaton field to gravity~\cite{GB2013,KL2013}. Note that quantum corrections to the action of the scalar field minimally coupled to gravity induce non-minimal coupling term~\cite{Chernikov_0, Chernikov_1}, proportional to $R\phi^2$ (where $R$ is the curvature and $\phi$ is the inflaton field).  Results of one-loop calculations in a weak gravitational field demonstrate ~\cite{Callan:1970ze,BirrellDavies,BOS} that in order to renormalize the theory
of a scalar field in curved space-time it is necessary to introduce
an induced gravity term proportional to $R\phi^2$. It follows that   one should add non-minimal coupling terms in the inflationary models to take into account quantum properties of the inflaton.

The idea of a supersymmetric (SUSY) inflationary model has been proposed in 1980s~\cite{ellis,SUSEinflation_0}. In particular, a number of advantages of simplified SUSY Grand Unified Theories (GUTs) in comparison with nonsupersymmetric GUTs such as a naturally longer period of exponential expansion and better stability of the effective Higgs potential with respect to radiative corrections due to a cancelation of loop diagrams has been noted in~\cite{ellis}.  The possibility to describe the inflation using particle physics models attracts a lot of attention (for a review see~\cite{Lyth:1998xn}) because it is considered as a fundamental step towards the unification of physics at all energy scales. In numerous models  the role of the inflaton has been performed by the Standard Model (SM) Higgs boson~\cite{Cervantes-Cota1995, higgsinf_0, higgsinf_1, higgsinf_2, higgsinf_3, higssinflRG_0, higssinflRG_1, higssinflRG_2,higgstop,KaiserHiggs} or a boson in GUTs~\cite{Barvinsky:1994hx,GUT_Inflation_0, GUT_Inflation_1} or a scalar boson in supersymmetric models~\cite{SUSEinflation_0, SUSEinflation_1, SUSEinflation_2,einhorn,Ferrara:2010yw}, see reviews~\cite{MazumdarRev,Ferrara:2015cwa}.

New particles consistent with restrictions on the new physics imposed by the LHC data provide extensive opportunities to improve significantly the Higgs-driven inflationary model. Multifield scenarios under consideration are based on a general observation that redefined fields in the Einstein frame  practically coincide with primary fields in the Jordan frame at the low energy scale of the order of superpartners mass scale $M_{SUSY}$, reproducing the two-Higgs doublet  potential, while at the scale higher than the GUT scale, the form of the potential in the redefined fields can be flat  to ensure the slow-roll approximation in inflationary scenarios.

In our paper~\cite{Dubinin:2017irg}, we have considered the model with two Higgs doublets coupled to gravity non-minimally. Convolutions of the Higgs doublets up to dimension-four in the fields form the well-known MSSM two-Higgs doublet potential which is rewritten in the mass basis of scalar fields~\cite{full_mssm}. This potential includes three massless Goldstone bosons and five massive Higgs bosons. Working in the physical gauge, when Goldstone bosons are not taken into account, we consider inflationary scenarios that include five Higgs bosons. It was shown~\cite{PlanckIfl_1} that inflationary scenarios with suitable parameters $n_\mathrm{s}$ and $r$ are possible at the scale corresponding to the Hubble parameter $H\sim 10^{-5}M_{\mathrm{Pl}}$. In~\cite{Dubinin:2017irg} we restrict ourselves to the special case when one of the vacuum expectation values is equal to zero, so,  the corresponding mixing angle $\beta=\pi/2$ (the so-called Higgs basis of scalar fields).  In the present work, it is shown that inflationary scenarios compatible with observational data are possible for other values of the mixing angle~$\beta$.
As a rule~\cite{giudice}, the limits on $\tan \beta$ coming from models of low-energy supersymmetry
are assumed to be $1< \tan \beta \leqslant m_{top}/m_b\simeq 35$, however
a rough lower bound corresponding to a Higgs boson--top quark coupling in the perturbative region can be of about $\tan \beta \geqslant m_{top}/600$ GeV \cite{hh93} and
the latest results of ATLAS and CMS collaborations~\cite{lhc_susy} show that
regions of large $\tan \beta$ at the 95\% confidence level (CL) should be excluded~\cite{large_tan}.

\section{The MSSM-inspired Higgs potential}\label{sec:2}

Let us consider the MSSM-inspired cosmological model which is described by the following action
\begin{equation}
\label{action}
    S=\int d^4x\sqrt{-g}[f(\Phi_1,\Phi_2)R-\delta^{ab}g^{\mu\nu}\partial_\mu\Phi_a^{\dagger}\partial_\nu\Phi_b
    -\, V(\Phi_1,\Phi_2)],
\end{equation}
where $\Phi_a$, $a=1,2$ are the Higgs doublets and the function $f$ is taken to be
\begin{equation}
\label{fp2}
    f(\Phi_1,\Phi_2)=\frac{M^2_{\mathrm{Pl}}}{2}+\xi_1\Phi_1^\dagger\Phi_1+\xi_2\Phi_2^\dagger\Phi_2\,.
\end{equation}
Here $M_\mathrm{Pl}\equiv1/\displaystyle\sqrt{8\pi G}$ is the reduced Planck mass, the constants $\xi_1$ and $\xi_2$ are positive and dimensionless.
Such form of the function $f$ follows from the requirement of renormalizability for quantum field theories in curved space-time~\cite{Chernikov_0, Chernikov_1, Callan:1970ze,BOS}, where non-minimal couplings appear as renormalization counterterms for scalar fields. We assume that vacuum expectation values for scalar fields are negligibly small in comparison with $M_{\mathrm{Pl}}$. Note that non-minimal interaction in the form of Eq.~(\ref{fp2}) has been considered in the framework of the general two-Higgs-doublet model with discrete $Z_2$ symmetry in~\cite{gong}.

The most general renormalizable two-doublet effective potential can be written as~\cite{hh93,Akhmetzyanova:2004cy_0, Akhmetzyanova:2004cy_1}:
\begin{equation}
\label{genV}
    V(\Phi_1,\Phi_2) =V_2(\Phi_1,\Phi_2)+V_4(\Phi_1,\Phi_2),
\end{equation}
where
\begin{equation}
V_2={}- \, \mu_1^2 (\Phi_1^\dagger\Phi_1) - \, \mu_2^2 (\Phi_2^\dagger
\Phi_2) - [ \mu_{12}^2 (\Phi_1^\dagger \Phi_2) +h.c.],
\end{equation}
\begin{eqnarray}
\label{V4}
V_4&=& \lambda_1
(\Phi_1^\dagger \Phi_1)^2
      +\lambda_2 (\Phi_2^\dagger \Phi_2)^2
+ \lambda_3 (\Phi_1^\dagger \Phi_1)(\Phi_2^\dagger \Phi_2) +
\lambda_4 (\Phi_1^\dagger \Phi_2)(\Phi_2^\dagger \Phi_1) \nonumber \\
&+& \left[\frac{\lambda_5}{2}
       (\Phi_1^\dagger \Phi_2)(\Phi_1^\dagger\Phi_2)+ \lambda_6
(\Phi^\dagger_1 \Phi_1)(\Phi^\dagger_1 \Phi_2)+\lambda_7 (\Phi^\dagger_2 \Phi_2)(\Phi^\dagger_1 \Phi_2)+h.c.\right].
\end{eqnarray}

Two Higgs doublets of the MSSM can be parameterized using the $SU(2)$ states
\begin{equation}
\label{dublets}
\Phi_1 = \left(\begin{array}{c} -\mathrm{i} \omega_1^+ \\ \frac{1}{\sqrt{2}} (v_1+\eta_1+\mathrm{i} \chi_1) \end{array} \right),\qquad
\Phi_2 = \left(\begin{array}{c} -\mathrm{i} \omega_2^+ \\ \frac{1}{\sqrt{2}} (v_2 +\eta_2+\mathrm{i} \chi_2) \end{array} \right),
\end{equation}
where $\omega^+_{1,2}$ are complex scalar fields, $\eta_{1,2}$ and $\chi_{1,2}$ are real scalar fields.

  The dimensionless factors $\lambda_i$ ($i=1,...,7$) at the tree level are expressed, using the $SU(2)$ and $U(1)$ gauge couplings $g_2$ and $g_1$, in the form~\cite{full_mssm_2, flores83}:
\begin{equation}
\lambda_1^{\tt tree}=\lambda_2^{\tt tree}=\frac{g_1^2+g_2^2}{8}, \quad
\lambda_3^{\tt tree}=\frac{g_2^2-g_1^2}{4},\quad
\lambda_4^{\tt tree}=-\frac{g_2^2}{2}, \quad
\lambda_{5,6,7}^{\tt tree}=0.
\end{equation}
At the superpartners mass scale $M_{SUSY}$ renormalization group evolved tree-level quartic couplings $\lambda_i$ can be evaluated using the collider data for $g_1$ and $g_2$ at $m_{top}$ scale. Indeed, the gauge boson masses at tree level $m_Z=v\sqrt{g^2_1+g^2_2}/2$, $m_W=v\, g_2/2$
($v=\sqrt{v^2_1+v^2_2}=(G_F\sqrt{2})^{-1/2}$, where $G_F$ is the Fermi constant). Substituting pole masses $m_Z=91.2$ GeV, $m_W=80.4$ GeV, and $v=246$ GeV we obtain $g_1=0.36$ and $g_2=0.65$ at the $m_{top}$ scale.

The mass basis of scalars is constructed in a standard way~\cite{full_mssm_2,Dubinin:2017irg}. The $SU(2)$ eigenstates $\omega^\pm_a$, $\eta_a$ and $\chi_a$ ($a=1,2$) are expressed through mass eigenstates of the Higgs bosons (two CP-even scalars $h$, $H_0$, pseudoscalar {$A$} and two charged bosons $H^\pm$) and the Goldstone bosons $G^0$, $G^\pm$  by means of two orthogonal rotations with angles $\alpha$ and $\beta$. In this paper, we use the unitary gauge $G^0=G^\pm=0$, therefore,
\begin{equation}
\label{equalpha}
\eta_1=\cos(\alpha)H_0-\sin(\alpha)h,\qquad \eta_2=\sin(\alpha)H_0+\cos(\alpha)h,
\end{equation}
\begin{equation}
\label{equbeta}
\chi_1={}-\sin(\beta)A, \quad \chi_2=\cos(\beta)A,\quad \omega_1^\pm={}-\sin(\beta)H^\pm,\quad \omega_2^\pm=\cos(\beta)H^\pm.
\end{equation}

In the following the $h$-boson is identified as the observed $125.09 \pm 0.24$ GeV scalar state~\cite{lhc-higgs}, so the 'alignment limit' of the MSSM is used where the  $h$-boson couplings to the gauge bosons and fermions are SM-like in agreement with the LHC data. In this limiting case $\beta-\alpha\approx \pi/2$.
Masses of Higgs bosons are negligibly small compared with the Planck mass. At tree level the mass $m_A$ and $t_\beta\equiv\tan \beta=v_2/v_1$ can be chosen as the input parameters which fix the dimension-two parameters $\mu^2_1,\mu^2_2$ and $\mu^2_{12}$ of the Higgs potential (see~\cite{Dubinin:2017irg} for detail). Insofar as during inflation the condition $V_4\gg V_2$ is respected, it is a good approximation to consider $V_4$ as the potential of the model neglecting the dimension-two convolutions. We use this approximation to get inflationary scenarios.

In the paper~\cite{Dubinin:2017irg}, we have considered the case of mixing angle $\beta = \pi/2$, then $\tan \beta$ is infinite. In this paper, the case of an arbitrary $\beta$ with a finite value of
$t_\beta$ is investigated.
Such a situation is typical for various MSSM scenarios which are used to describe the LHC data. If the radiative corrections are included,
a number of such effective field theories with the boundary conditions
fixed at $m_{top}$ scale have been considered. A discussion of 'benchmark
scenarios' which are denominated as $m^{max}_h$, $m^{mod+}_h$,
$m^{mod-}_h$, {\it light stop}, {\it light stau} and $\tau$-{\it phobic}
and restrictions on the MSSM parameter space consistent with the LHC
data can be found in \cite{benchmark}. Specific case of parametric
scenarios where $\tan \beta \sim 1$ is analysed in~\cite{hxswg}.

Assuming the alignment limit with $\beta = \pi/2+\alpha$, when the branching ratios of the observed resonance with the mass at $125$ GeV are SM-like, and using Eqs.~(\ref{equalpha}) and (\ref{equbeta}), we obtain
\begin{equation}\label{Phi1Phi1tg}
(\Phi_1^\dag\Phi_1)=\frac{1}{2(t_\beta^2+1)}\left[(A^2+H_0^2+2H^-H^+)t_\beta^2+2H_0h_vt_\beta+h_v^2\right],
\end{equation}
\begin{equation}\label{Phi2Phi2tg}
(\Phi_2^\dag\Phi_2)=\frac{1}{2(t_\beta^2+1)}\left[h_v^2t_\beta^2-2H_0h_vt_\beta+A^2+H_0^2+2H^-H^+\right],
\end{equation}
\begin{equation}\label{Phi1Phi2tg}
(\Phi_1^\dag\Phi_2)=\frac{1}{2(t_\beta^2+1)}\left[h_v(H_0+\mathrm{i}A)t_\beta^2+(h_v^2-A^2-H_0^2-2H^-H^+)t_\beta+(\mathrm{i}A-H_0)h_v\right],
\end{equation}
where $h_v=h+v$.

Using Eqs.~(\ref{Phi1Phi1tg})--(\ref{Phi1Phi2tg}), the potential~$V_4$ can be written in terms of the scalar fields:
\begin{eqnarray*}
V_4 &=& \frac{1}{32\left(t_\beta^2+1\right)^2} \left\{\left(g_1^2+g_2^2\right)\left(t_\beta^2-1\right)^2 h_v^4-8\left(g_1^2+g_2^2\right)t_\beta\left(t_\beta^2-1\right)H_0h_v^3
 \right.
\\
&+& \left[\left(2(2H^-H^+ -A^2-H_0^2)g_2^2-2(A^2+H_0^2+2H^-H^+)g_1^2\right)t_\beta^4\right.\\
&+&\left((4A^2+20H_0^2+8H^-H^+)g_1^2+4(A^2+5H_0^2+6H^-H^+)g_2^2\right)t_\beta^2\\
&+&\left. 2(2H^-H^+-A^2-H_0^2)g_2^2 -2(A^2+H_0^2+2H^-H^+)g_1^2\right]h_v^2 \\
&+& 8(g_1^2+g_2^2)t_\beta\left(t_\beta^2-1\right)(A^2+H_0^2+2H^-H^+)H_0h_v \\
&+&\left. \left(g_1^2+g_2^2\right)\left(t_\beta^2-1\right)^2(A^2+H_0^2+2H^-H^+)^2\right\}.
\end{eqnarray*}

The function $f$ in terms of the scalar fields can be written as
\begin{equation}\label{f_fields}
\begin{split}
     f&=\frac{M^2_{\mathrm{Pl}}}{2}+\frac{\xi_1}{2(t_\beta^2+1)}\left[(A^2+H_0^2+2H^-H^+)t_\beta^2+2H_0h_vt_\beta+h_v^2\right]\\&{}+
    \frac{\xi_2}{2(t_\beta^2+1)}\left[h_v^2t_\beta^2-2H_0h_vt_\beta+A^2+H_0^2+2H^-H^+\right].
\end{split}
\end{equation}

\section{Cosmological attractors}
\label{sec:3}

In this section, we show how the idea of cosmological attractors~\cite{Roest:2013fha, LindeKallosh_etal_0, Kaiser:2013sna, Kallosh:2013daa, Binetruy:2014zya_0, Ventury2015, EOPV2016,Odintsov:2016vzz,string_motivated} which is based on an observation that the kinetic term in the Jordan frame practically does not affect the slow-roll parameters (so-called  'strong coupling regime')  allows to get approximate values of the inflationary parameters  $n_\mathrm{s}$ and $r$ in the analytic form.

The potential $V$ depends on five real scalar fields
\begin{equation}
\phi^1=\frac{H^+ +H^-}{\sqrt{2}},\, \phi^2=\frac{H^+ -H^-}{\sqrt{2}\mathrm{i}},\, \phi^3=A, \, \phi^4=H_0,\, \phi^5=h_v.
\end{equation}
The action defined by Eq.~(\ref{action}) with $f$ and $V$ written in terms of the scalar fields $\phi^I$ can be transformed to the following action in the Einstein frame~\cite{Kaiser:2010ps} (see also~\cite{KaiserHiggs,Kaiser}):
\begin{equation}
\label{SE}
S_{E}=\int d^4x\sqrt{-g}\left[\frac{M^2_{\mathrm{Pl}}}{2}R-\frac12\mathcal{G}_{IJ}{g^{\mu\nu}}\partial_\mu\phi^I\partial_\nu\phi^J-W\right],
\end{equation}
where
\begin{equation*}
\mathcal{G}_{IJ}=\frac{M^2_{\mathrm{Pl}}}{2f(\phi^K)}\left[\delta_{IJ}
+\frac{3f_{, I} f_{, J}}{f(\phi^K)}\right],\qquad W= M^4_{Pl}\frac{V}{4f^2},
\end{equation*}
$f_{,I} = \partial f/\partial \phi^I$. Metric tensors
in the Jordan and in the Einstein frames are related by the equation
\begin{equation*}
g_{\mu\nu}=\frac{2f(\phi^I)}{M^2_{\mathrm{Pl}}}\tilde{g}_{\mu\nu}.
\end{equation*}

By definition the strong coupling regime of the field system takes place if the following inequality is respected
\begin{equation}
\label{StrCC}
\delta_{IJ}\partial_\mu\phi^I\partial_\nu\phi^J\ll\frac{3}{f(\phi^K)}f_{,I}f_{,J}\partial_\mu\phi^I\partial_\nu\phi^J.
\end{equation}
Using Eq.~(\ref{SE}) in this approximation, we get
\begin{equation}
\label{SEapr}
S_{E}
=
\!\int\! d^4x\sqrt{-g}\left[\frac{M^2_{\mathrm{Pl}}}{2}R-\frac{g^{\mu\nu}}{2}\partial_\mu \Theta
\partial_\nu \Theta - \frac{M^4_{Pl}V_4}{4f^2}\right],
\end{equation}
where
\begin{equation}
\label{lnf}
    \Theta=\sqrt{\frac32}M_{\mathrm{Pl}}\ln\left(\frac{f}{f_0}\right)
\end{equation}
and
$f_0$ is a positive constant with the same dimension as $f$. If we choose $f_0={M^2_{\mathrm{Pl}}}/{2}$, then $\Theta=0$ corresponds to zero values of all scalar fields. We assume that $h_v$ is marginal and consider the inflationary scenario in the case $\phi^5=h_v=0$ during inflation. Note that in~\cite{Dubinin:2017irg} this scenario has been considered at $\beta=\pi/2$ (denominated in~\cite{Dubinin:2017irg} as the 'case $A$') which gives suitable inflationary parameters. One can observe that in the strong coupling regime the system can be described by the reduced action with only one scalar field $\Theta$. Indeed, from~(\ref{f_fields}) we get
\begin{equation}
\label{fappr}
    f=\frac{M^2_{\mathrm{Pl}}}{2}+\frac{(\xi_1t_\beta^2+\xi_2)(A^2+H_0^2+2H^+H^-)}{2\left(t_\beta^2+1\right)}.
\end{equation}
It follows that the potential $V_4$ can be expressed as a function of $f$:
\begin{equation}
\label{Vappr}
    V_4= \frac{1}{32}  (g_1^2+g_2^2)\left(t_\beta^2-1\right)^2\left(\frac{2f-M^2_{\mathrm{Pl}}}{\xi_1t_\beta^2+\xi_2}\right)^2.
\end{equation}
In the Einstein frame the potential is
\begin{equation}
\label{Wappr}
W=\frac{M^4_{Pl}V_4}{4f^2}=\frac{\left(g_1^2+g_2^2\right)M^4_{Pl}}{32\left(\xi_1t_\beta^2+\xi_2\right)^2}\left(t_\beta^2-1\right)^2
\left(1-e^{-\sqrt{\frac23}\frac{\Theta}{M_{\mathrm{Pl}}}}\right)^2.
\end{equation}

The slow-roll approximation from a simple single-field model can be used to get inflationary parameters.
In the spatially flat Friedman--Lemaitre--Robertson--Walker (FLRW) universe with the interval
\begin{equation*}
ds^2={}-dt^2+a^2(t)\left(dx_1^2+dx_2^2+dx_3^2\right),
\end{equation*}
where $a(t)$ is the scale factor, the slow-roll parameters are

\begin{equation*}
\epsilon = \frac{M^2_{\mathrm{Pl}}}{2}\left(\frac{W^\prime_{\Theta}}{W}\right)^2=\frac{4\;\mathrm{e}^{-{\frac {2\sqrt{6}\Theta}{3M_{\mathrm{Pl}}}}}}{3 \left( 1-\mathrm{e}^{-{\frac {\sqrt {6}\Theta}{3M_{\mathrm{Pl}}}}} \right)^2},
\end{equation*}
\begin{equation*}
  \eta=M_\mathrm{Pl}^{2}\frac{W^{\prime\prime}_{\Theta\Theta}}{W}=\frac{4\mathrm{e}^{-{\frac {\sqrt{6}\Theta}{3M_{\mathrm{Pl}}}}} \left( 2\mathrm{e}^{-{\frac {\sqrt{6}\Theta}{3M_{\mathrm{Pl}}}}}-1 \right)}{ 3 \left(1-
\mathrm{e}^{-{\frac {\sqrt{6}\Theta}{3M_{\mathrm{Pl}}}}}\right)^2}=\frac{4 \left(2- \mathrm{e}^{{\frac {\sqrt{6}\Theta}{3M_{\mathrm{Pl}}}}} \right)}{ 3 \left(1-
\mathrm{e}^{{\frac {\sqrt{6}\Theta}{3M_{\mathrm{Pl}}}}}\right)^2},
\end{equation*}
where primes denote derivatives with respect to $\Theta$.

It is convenient to express the inflationary parameters
\begin{equation}\label{nsf}
   n_\mathrm{s}=1-6\epsilon+2\eta=1- \frac{8\;\mathrm{e}^{-\frac {\sqrt{6}\Theta}{3M_{\mathrm{Pl}}}} \left(
1+\mathrm{e}^{-{\frac {\sqrt{6}\Theta}{3M_{\mathrm{Pl}}}}} \right) }{3\left(
1-\mathrm{e}^{-{\frac {\sqrt{6}\Theta}{3M_{\mathrm{Pl}}}}} \right)^2},
\end{equation}
\begin{equation}
\label{rf}
r=16\epsilon=\frac{64\;\mathrm{e}^{-{\frac {2\sqrt{6}\Theta}{3M_{\mathrm{Pl}}}}}}{3 \left( 1-\mathrm{e}^{-{\frac {\sqrt {6}\Theta}{3M_{\mathrm{Pl}}}}} \right)^2}.
\end{equation}

Using Eq.~(\ref{f_fields}), we get
  \begin{equation}\label{nsrf}
   n_\mathrm{s}=1-\frac{8M^2_{\mathrm{Pl}}\left(M^2_{\mathrm{Pl}}+2f\right)}{3\left(M^2_{\mathrm{Pl}}-2f\right)^2},\qquad r=\frac{64M^4_{\mathrm{Pl}}}{3\left(M^2_{\mathrm{Pl}}-2f\right)^2}\,.
\end{equation}

Note that these expressions for $n_\mathrm{s}(f)$ and $r(f)$ do not depend on $t_\beta$ and coincide with the corresponding formulae for $\beta = \pi/2$ (when $t_\beta = \infty$) which has been obtained in~\cite{Dubinin:2017irg}. At the same time it is not correct to say that the corresponding inflationary scenarios do not depend on $\beta$, the direction to potential minima in the $(v_1,\,v_2)$-plane, because the Hubble parameter is expressed through~$f$ as follows:
\begin{equation}
H_f^2\approx{\frac {M_{\mathrm{Pl}}^{2} \left( {{g_1}}^{2}+{{g_2}}^{2} \right)  \left(t_\beta^{2}-1 \right) ^{2} \left( M_{\mathrm{Pl}}^{2}-2f \right)^{2}}{
 384 \left(\xi_1\,t_\beta^{2}+\xi_2 \right)^2 {f}^{2}}}.
\end{equation}
One can see that inflationary scenarios  are excluded at $\beta=\pi/4$.

The temperature data of the Planck full mission and first release of the polarization data at large angular scales~\cite{PlanckIfl_1} constrain the
spectral index of curvature perturbations to
\begin{equation}\label{nsdata}
    n_\mathrm{s}=0.968 \pm 0.006 \quad (68\%\, {\rm CL}),
\end{equation}
and restrict the tensor-to-scalar ratio from above
\begin{equation}\label{rdata}
    r < 0.11 \quad (95\%\, {\rm CL}).
\end{equation}
Using the observable value of $n_\mathrm{s}=0.968$, we obtain from Eq.~(\ref{nsrf}) the corresponding value of $f=43.14M_{\mathrm{Pl}}^2$, so the Hubble parameter is expressed in a compact form
\begin{equation}
\label{Hf}
H_{f}^2\approx  0.01\,{\frac { \left( t_\beta^{2}-1 \right) ^{2}M_{\mathrm{Pl}}^{2}
 \left( g_1^{2}+g_2^{2} \right) }{\left(\xi_1\,t_\beta^{2}+\xi_2 \right)^2}}.
\end{equation}

As mentioned in the Introduction, there is an upper bound on the Hubble parameter during inflation~\cite{PlanckIfl_1}: $H < 3.6 \times 10^{-5}M_{\mathrm{Pl}}$.
Using Eq.~(\ref{Hf}) with given values of the model parameters, one can easily check whether this upper bound is respected. More precise numerical calculations are essentially simplified when suitable initial values of $f$ can be easily found in the strong coupling approximation.

\section{Numerical calculations in the Einstein frame formulation}
\label{sec:4}

Varying the Einstein frame action~(\ref{SE}) with respect to $g_{\mu \nu}$ and fields, one can get the following equations for the spatially flat FLRW metric~\cite{Kaiser}:
\begin{equation}
H^2= \frac{1}{3M_{\mathrm{Pl}}^2}\left(\frac{\dot{\sigma}^2}{2}+W\right), \qquad
  \dot{H}={} -\frac{1}{2M_{\mathrm{Pl}}^2}\dot{\sigma}^2, \label{Equations}
\end{equation}
\begin{equation}
\label{EqFields}
\ddot\phi^I+3H\dot\phi^I + \Gamma^I_{\>\> JK}  \dot \phi^J \dot \phi^K + \mathcal{G}^{IK} W'_{, K} = 0\,,
\end{equation}
where $\Gamma^I_{\>\> JK}$ is the Christoffel symbol for the field-space manifold with the metric $\mathcal{G}_{IJ}$,
 $W{'}_{,K} = \partial W/\partial \phi^K$ and
$\dot{\sigma}^2=\mathcal{G}_{IJ} \dot \phi^I \dot \phi^J.$

During inflation the scalar factor $a$ is a monotonically increasing function. So, we can use the number of e-foldings $N_e=\ln(a/a_{e})$, where $a_e$ is the value of the scalar factor at the end of inflation, as a new measure of time. Due to the relation ${d}/{dt}=H\, {d}/{dN_e}$, Eqs.~(\ref{Equations}) and (\ref{EqFields}) can be written in the following form~\cite{Dubinin:2017irg}:
\begin{equation}
  H^2=\frac{2W}{6M_{\mathrm{Pl}}^2-(\sigma^\prime)^2}, \qquad \frac{d\ln{H}}{dN_e}={} -\frac{1}{2M_{\mathrm{Pl}}^2}\left(\sigma^\prime\right)^2, \label{H2NdlnHN}
\end{equation}
\begin{equation}
  \frac{d\phi^I}{dN_e}=\psi^I,\qquad
  \frac{d\psi^I}{dN_e}={} -\left(3+\frac{d\ln{H}}{dN_e}\right)\psi^I - \Gamma^I_{\>\> JK}\psi^J\psi^K -  	  \frac{1}{H^2}\mathcal{G}^{IK} W^{'}_{,K} , \label{diff eqN}
\end{equation}
where $(\sigma')^2 = H^2 \, (\dot{\sigma})^2$.
So, we get the following system of ten first order equations
\begin{equation}
\label{systemN}
    \left\{
\begin{split}
 \frac{d\phi^I}{dN_e}&=\psi^I,\\
 \frac{d\psi^I}{dN_e}&={} -\left(3-\frac{\left(\sigma^\prime\right)^2}{2M_{\mathrm{Pl}}^2}\right)\psi^I - \Gamma^I_{\>\> JK}\psi^J\psi^K - \frac{6M_{\mathrm{Pl}}^2-(\sigma^\prime)^2}{2W}\mathcal{G}^{IK} W^{'}_{,K}.
\end{split}
    \right.
\end{equation}

In order to calculate the spectral index $n_\mathrm{s}$ and the tensor-to-scalar ratio $r$, the slow-roll parameters are introduced analogously to the single-field inflation
\begin{equation}
\epsilon = {}-\frac{\dot{H}}{H^2},\qquad
\eta_{\sigma\sigma} = M_{\mathrm{Pl}}^2 \frac{{\cal M}_{\sigma\sigma}}{W},
\qquad
{\cal M}_{\sigma\sigma} \equiv \hat{\sigma}^K \hat{\sigma}^J ({\cal D}_K {\cal D}_J W),
\end{equation}
where $\sigma^I=\dot{\phi^I}/\dot{\sigma}$ is the unit vector in the field space and $\cal D$ denotes a covariant derivative with respect to the field-space metric: ${\cal D}_I\phi^J=\partial_I \phi^J+ \Gamma^J_{IK}\phi^K$.

In the case of mixing angle $\beta=\pi/2$ suitable inflationary scenarios have been found in~\cite{Dubinin:2017irg}, where various combinations of scalar fields $\phi^1$,..., $\phi^5$ play the role of inflaton. The goal of this paper is to show that inflationary scenarios for finite values of $t_\beta$ are also possible and to find the corresponding values of inflationary parameters.
\begin{figure}
\centering
\includegraphics[width=0.47\linewidth]{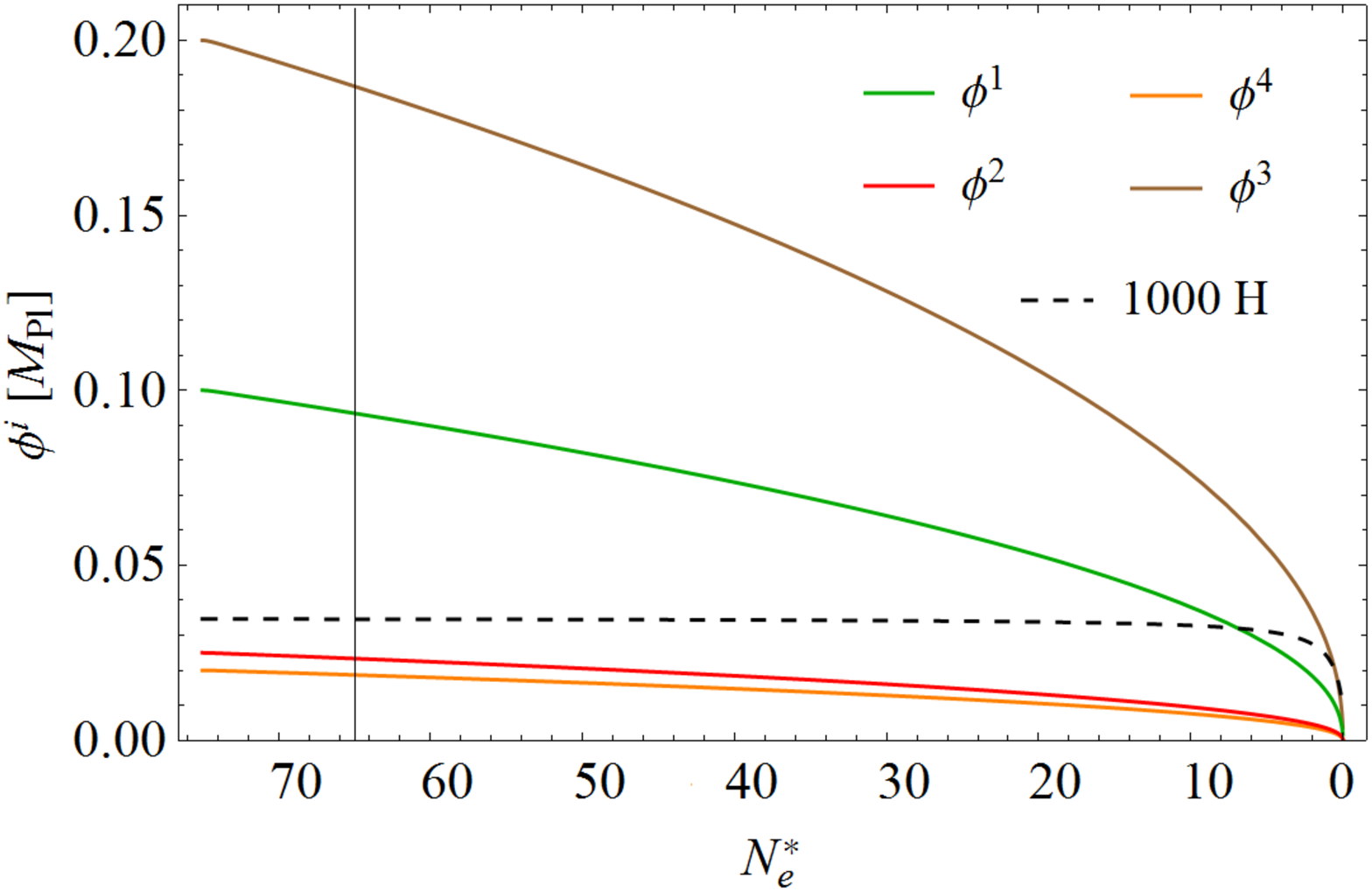}\ \ \ \
\includegraphics[width=0.47\linewidth]{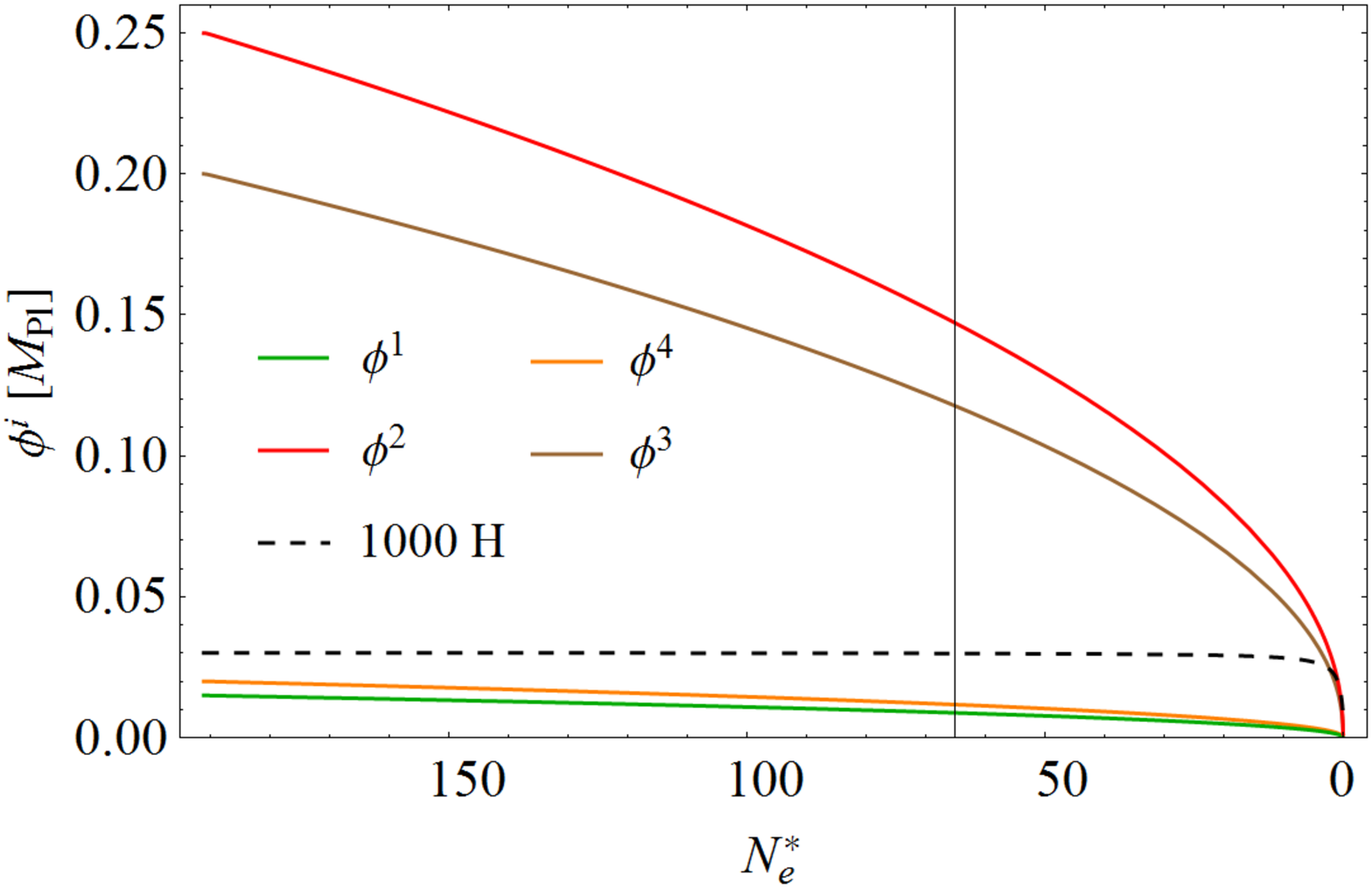}
\caption{Evolution of the fields and the Hubble parameter as functions of the number of e-foldings $N_e^*=-N_e$ during inflation in scenarios $A_1$ (left) and $C_1$ (right), see Table~1. The vertical line corresponds to $N_e^*=65$.}
\label{Fig1}
\end{figure}

As mentioned above, the approximation $V\simeq V_4$ is used at $h_v \approx 0$. In this case, the formulae obtained in the strong coupling approximation are validated.
Numerical calculations show (Fig.~\ref{Fig1}) that all scalar fields monotonically decrease before and during inflation, so the function $f$ is also a monotonically decreasing function.  For this reason, such initial values of scalar fields are chosen that the corresponding values of $f$ are greater than $43.14M_{\mathrm{Pl}}^2$. Thus, the strong coupling approximation simplifies the choice of initial conditions for numerical calculations. The list of initial conditions giving acceptable inflationary scenarios is presented in Table~1.

\begin{table}[ht]
\label{Tabl1}
{\begin{tabular}{|c|c|cc|cccc|}
\hline
Scenario & $t_\beta $ & $\xi_1$ & $\xi_2$ & $\phi_1/M_{\mathrm{Pl}}$ & $\phi_2/M_{\mathrm{Pl}}$ & $\phi_3/M_{\mathrm{Pl}}$ & $\phi_4/M_{\mathrm{Pl}}$ \\
\hline
$A_1$ & 5 & 2000 & 2000 & $0.1$ & $0.025$ & $0.2$& $0.02$\\
$A_2$ & 5 & 2000 & 2000 & $0.2$ & $0.05$ & $0.1$ & $0.15$\\
$B_1$ & 10 & 2500 & 2500 & $0.15$ & $0.15$ & $0.1$ & $0.1$\\
$B_2$ & 10 & 2000 & 1000 & $0.15$ & $0.15$ & $0.1$ & $0.2$\\
$C_1$ & 20 & 2500 & 1000 & $0.015$ & $0.25$ & $0.2$ & $0.02$\\
$C_2$ & 20 & 2500 & 500 & $0.015$ & $0.25$ & $0.2$ & $0.02$\\
$D_1$ & 40 & 2000 & 2000 & $0.01$ & $0.025$ & $0.2$ & $0.02$\\
$D_2$ & 40 & 2000 & 2000 & $0.12$ & $0.12$ & $0.12$ & $0.12$\\
\hline
\end{tabular}}
\caption{The parameters of the model and the initial field values for numerical calculations.}
\end{table}

The spectral index $n_\mathrm{s}$ and tensor-to-scalar ratio $r$ at the time when a characteristic scale  is of the order of the Hubble radius (50--65 e-foldings before the end of inflation) can be expressed via the slow-roll parameters~\cite{Kaiser,c_perturbations}
\begin{equation}
n_\mathrm{s}=1-6 \epsilon+2 \eta_{\sigma \sigma}, \qquad
r=16 \epsilon.
\label{nsr_numeric}
\end{equation}

 In Table~2 
the corresponding values of function $f$ at $N_e=-65$, when inflationary parameters are calculated, are presented. One can observe that in all inflationary scenarios results of numerical calculations of the inflationary parameters are close to their values evaluated using symbolic formulae obtained in the strong coupling approximation (see Table~2).

\begin{table}[ht]
{\begin{tabular}{|c|c|c|c|c|}
\hline
Scenario &$f/M_{\mathrm{Pl}}^2$ & $H/M_{\mathrm{Pl}}$
[$10^{-5}$] & $r$ & $n_\mathrm{s}$\\  \hline
$A_1$  & 45.042 & 3.461 &0.002661 & 0.9694 \\
$A_2$ & 45.247 & 3.462 & 0.002637 & 0.9695 \\
$B_1$ & 44.807 & 2.941 & 0.002649 & 0.9695 \\
$B_2$ & 45.048& 3.694 & 0.002660 & 0.9694 \\
$C_1$& 44.910 & 2.989 & 0.002677 & 0.9695\\
$C_2$& 45.084 & 2.991 & 0.002656 & 0.9694 \\
$D_1$& 44.908 &3.461& 0.002661 & 0.9694 \\
$D_2$& 45.387 &3.462& 0.002621 & 0.9696 \\
\hline
\end{tabular}}
\caption{The values of function $f$ and the Hubble parameter $H$ at $N_e=-65$, together with the tensor-to-scalar ratio $r$ and spectral index $n_\mathrm{s}$, obtained by Eq.~(\ref{nsr_numeric}), for successful inflationary scenarios.}
\label{Tabl2}
\end{table}

\section{Summary}\label{sum}
In this paper, a MSSM-inspired extension of the original Higgs-driven inflation~\cite{higgsinf_0, higgsinf_1, higgsinf_2, higgsinf_3, higssinflRG_0, higssinflRG_1, higssinflRG_2} is constructed using the two-Higgs doublet potential, given by Eq.~(\ref{genV}). During inflation the quadratic part of this potential is negligibly small, so the potential can be approximated by its fourth order part $V_4$. Assuming that the field $h_v$ is negligibly small during inflation, we simplify the potential in a way suitable for calculation of transparent symbolic and numerical results for the main observables: the spectral index~$n_\mathrm{s}$ and the tensor-to-scalar ratio~$r$. The inflationary scenarios under consideration incorporate four non-minimally coupled scalar fields. It is shown that the considered model  can be mapped to the single-field model with the effective inflaton field defined by Eq.~(\ref{lnf}) using the strong coupling approximation. In this approximation the inflationary parameters do not depend on the mixing angle~$\beta$. On the other hand, numerical calculations in the strong coupling regime  are realized in an inflationary scenarios with different values of model parameters and initial conditions. Such models share very close results for the spectral index and the tensor-to-scalar ratio in combination with negligible non-Gaussianity, which are in good agreement with the latest experimental data. A generalization of the  MSSM-inspired  model analysed in~\cite{Dubinin:2017irg} for the case of finite values of $\tan \beta$ is found.

In conclusion, we should note that an important point beyond our analysis is the stability of results with respect to radiative corrections and the renormalization group evolution of $\xi_i$. The general case of renormalization group
improved (RG-improved) effective potential for gauge theories in curved
spacetime when a generalization of Coleman--Weinberg resummation for the
flat space is introduced can be found in~\cite{elizalde}, where several
applications to explicit field theory models have been presented.
A number of simple supersymmetric models with running coupling of scalar
field to gravity have been analysed. The simplest Wess--Zumino model
effective potential along a flat direction in de Sitter space~\cite{futamase} demonstrates the power-like running of $\xi$
typical for $\lambda \phi^4$ field theory~\cite{buchbinder}. The RG-improved effective
action in curved space coupled with the classical gravity includes the
term $\xi(\tilde{t}) \, R \phi^2$ which is analogous to the representation of
Eq.~(\ref{fp2}), $\tilde{t}\equiv \log(\phi^2/\mu^2)$, where $\mu$ is the renormalization scale. Inflation
with different evolution of $\xi$ defined by exponent $\exp(c\, g^2 \,
\tilde{t})$ has been observed in supersymmetric finite GUTs~\cite{odintsov},
where $g$ is the gauge coupling in the matter sector and the numerical
constant $c$ defined by the symmetry group of the grand unification
theory can be negative, positive or zero.
The initial conditions for the inflaton field which are
discussed in~\cite{futamase,odintsov} respect definite
restrictions (for example, $|c| \, g^2 \sim$ 10$^{-3}$ in the $SU(2)$
gauge theory with $SU(N)$ global invariance considered in~\cite{odintsov}) in order to have successful inflationary scenarios. So,
supersymmetric finite GUTs in curved spacetime are recognized as
interesting candidates in connection with the inflationary universe
scenarios and possible applications to the cosmological constant problem.
  The renormalization group corrections have been analyzed in different single-field inflationary scenarios motivated by different nonsupersymmetric models~\cite{higssinflRG_0, higssinflRG_1, higssinflRG_2, rgingl_0, rgingl_1, rgingl_2}. For a slow-roll it is essential to have nearly flat effective potential in the region of the field amplitudes of the order of $M_{\mathrm{Pl}}$.  While the quantum gravity corrections are expected to be not important, the corrections induced by the SM fields and the superpartner fields (see, for example, \cite{last} and references therein) require careful analysis which is dependent on the MSSM parametric scenario under consideration.

\section*{Acknowledgments}

 The authors are thankful to M.V.~Sumin and I.P.~Volobuev for useful discussions.
This work was partially supported by Grant No. NSh-7989.2016.2 of the President of the Russian Federation.
The research of E.O.P. and E.Yu.P. was supported in part by Grant No. MK-7835.2016.2 of the President of the Russian Federation.

\end{document}